\documentclass[11pt]{article}
\usepackage[totalwidth=460pt,totalheight=600pt]{geometry}
\usepackage{epsfig,latexsym}

\def\theequation{\arabic{section}.\arabic{equation}}

\renewcommand{\theequation}{\thesection.\arabic{equation}}
\global\arraycolsep=1truept
\input epsf
\input{tcilatex}
\begin{document}

\font\cmss=cmss10 \font\cmsss=cmss10 at 7pt \hfill \hfill IFUP-TH/03-38

\vskip 2truecm

\vspace{10pt}

\begin{center}
{\Large \textbf{\vspace{10pt}RENORMALIZATION\ OF\ QUANTUM\ GRAVITY COUPLED\
WITH\ MATTER\ IN\ THREE\ DIMENSIONS}}

\bigskip \bigskip

\textsl{Damiano Anselmi}

\textit{Dipartimento di Fisica ``E. Fermi'', Universit\`{a} di Pisa, and INFN%
}
\end{center}

\vskip 2truecm

\begin{center}
\textbf{Abstract}
\end{center}

{\small In three spacetime dimensions, where no graviton propagates, pure
gravity is known to be finite. It is natural to inquire whether finiteness
survives the coupling with matter. Standard arguments ensure that there
exists a subtraction scheme where no Lorentz-Chern-Simons term is generated
by radiative corrections, but are not sufficiently powerful to ensure
finiteness. Therefore, it is necessary to perform an explicit (two-loop)
computation in a specific model. I consider quantum gravity coupled with
Chern-Simons $U(1)$ gauge theory and massless fermions and show that
renormalization originates four-fermion divergent vertices at the second
loop order. I\ conclude that quantum gravity coupled with matter, as it
stands, is not finite in three spacetime dimensions.}

\vskip 1truecm

\vfill\eject

\section{Introduction}

\setcounter{equation}{0}

Gravity is not power-counting renormalizable in dimensions greater than two.
It is known \cite{thooftveltman} that pure gravity in four-dimensions is
finite to the first loop order and that one-loop finiteness is spoiled by
the coupling with matter. Moreorer, four-dimensional gravity is not finite
to the second loop order \cite{sagnotti}, even in the absence of matter.

In three dimensions there is no propagating graviton and pure gravity 
\begin{equation}
\frac{1}{2\kappa }\int \sqrt{g}R(x)~\mathrm{d}^{3}x  \label{3dG}
\end{equation}
is known to be finite to all orders \cite{witten}. A quick proof is based on
the observation that the counterterms vanish using the field equations of (%
\ref{3dG}) and therefore can be reabsorbed by means of field redefinitions.
Indeed, in three dimensions the Weyl tensor is identically zero and so the
Riemann tensor is a linear combination of the Ricci tensor and the scalar
curvature: 
\begin{equation}
R_{\mu \nu \rho \sigma }=g_{\mu \rho }R_{\nu \sigma }-g_{\mu \sigma }R_{\nu
\rho }-g_{\nu \rho }R_{\mu \sigma }+g_{\nu \sigma }R_{\mu \rho }-\frac{R}{2}%
g_{\mu \rho }g_{\nu \sigma }+\frac{R}{2}g_{\mu \sigma }g_{\nu \rho }.
\label{ricci}
\end{equation}
Every counterterm is proportional to $R_{\mu \nu }$ or $R$, apart from the
Lorentz-Chern-Simons term 
\begin{equation}
\int \varepsilon ^{\mu \nu \rho }\left( \omega _{\mu }^{a}\partial _{\nu
}\omega _{\rho }^{a}+\frac{1}{3}\omega _{\mu }^{a}\omega _{\nu }^{b}\omega
_{\rho }^{c}\varepsilon ^{abc}\right) ,  \label{gravChS}
\end{equation}
which does not appear by parity invariance. By dimensional counting, the
counterterms are actually quadratic, at least, in $R_{\mu \nu }$-$R$ and
therefore can be reabsorbed by means of covariant field redefinitions, with
no renormalization of the Newton constant $\kappa $.

It is natural to inquire whether finiteness survives the coupling with
matter in three dimensions. The renormalization of the theory has chances to
be non-trivial, even if no graviton propagates. If the theory is finite,
renormalization requires only field redefinitions, but no running of the
coupling constants. If the theory is not finite, then renormalization
generates infinitely many new coupling constants, as in four dimensions. In
this paper I study these issues.

First I analyze non-renormalization properties and standard arguments about
finiteness. I prove that there exists a subtraction scheme where no
Lorentz-Chern-Simons term is generated by radiative corrections. This
ensures that gravity is not driven to the theory known as ``topologically
massive gravity'' \cite{jackiw}. However, the standard non-renormalization
arguments are not sufficiently powerful to ensure finiteness, because
higher-dimensioned operators can be generated by renormalization. To decide
whether finiteness survives the coupling with matter or not, it is necessary
to perform an explicit computation in a specific model. I consider quantum
gravity coupled with Chern-Simons $U(1)$ gauge theory and massless fermions.
This model is a good laboratory to explore ideas about finiteness and
renormalizability beyond power counting. I show that renormalization
originates a four-fermion divergent vertex 
\begin{equation}
\mathcal{L}_{\text{div}}=\frac{5\kappa g^{4}n_{f}}{384\pi ^{2}\varepsilon }%
\frac{e}{4}(\overline{\psi }\gamma ^{a}\psi )^{2}  \label{divgrav}
\end{equation}
at the second order in the loop expansion and first order in the Newton
constant $\kappa $. The result (\ref{divgrav}) is written up to subleading
corrections in $1/n_{f}$. I\ conclude that quantum gravity coupled with
matter, as it stands, is not finite in three spacetime dimensions.

The computation is two-loop, because in three dimensions every theory is
finite to the first loop order. By symmetric integration, an odd-dimensional
theory has no one-loop logarithmic divergence. Moreover, the power-like
divergences are scheme artifacts (they are automatically absent using the
dimensional-regularization technique) and have no effects on the
renormalization group. So, the problem of finiteness starts at two loops in
three dimensions.

\bigskip

At the classical level, the properties of gravity coupled with matter in
three dimensions have been widely studied, starting from ref. \cite{jackiw2}%
. At the quantum level, there have been studies on quantum gravity of point
particles \cite{thooft}, quantum cosmology and black-hole quantum mechanics 
\cite{carlip}, topologically massive gravity \cite{jackiw}, gravitating
topological matter \cite{carlip2}, de Sitter quantum gravity \cite{regge},
loop quantum gravity \cite{loops}, dynamically triangulated quantum gravity 
\cite{amb1} and many other subjects.

In flat space, the renormalization of 2+1 dimensional quantum field theory
has been studied at the perturbative level \cite{avdeev,avdeev2,nonre} and
in the large N expansion \cite{parisi,gross,rosenstein,largeN,largeN2}.
Besides the finiteness of pure gravity in three dimensions \cite{witten},
there have been studies on the renormalizability of quantum gravity near two
dimensions \cite{giappo}. The renormalization of 2+1 quantum gravity coupled
with matter has attracted less attention, so far. The interest of this
research is that it can shed some light on the properties of renormalization
beyond power-counting.

The paper is organized as follows. In section 2 I\ recall the properties of
Chern-Simons $U(1)$ gauge theory with matter in flat space. In section 3 I
couple it with gravity. In section 4 I prove that no Lorentz-Chern-Simons
term is induced by renormalization. In section 5 I introduce the two-loop
computations of this paper, the organization of counterterms and the
calculational technique. In section 6 I collect the results about
four-fermion vertices induced by gravity. Section 7 contains the
conclusions. In the appendix I collect useful formulas and some remarks
about the difficulties of the dimensional regularization of the Chern-Simons
term.

\section{Chern-Simons $U(1)$ gauge theory with massless fermions}

\setcounter{equation}{0}

In this section I recall some properties of Chern-Simons $U(1)$ gauge theory
with massless fermions in flat space and fix the notation. I work in the
Euclidean framework. The lagrangian reads 
\begin{equation}
\mathcal{L}_{\mathrm{cl}}=\overline{\psi }D\!\!\!\!\slash\psi +\frac{1}{%
2g^{2}}\varepsilon ^{\mu \nu \rho }F_{\mu \nu }A_{\rho },  \label{conf}
\end{equation}
where $D_{\mu }=\partial _{\mu }+iA_{\mu }$ is the covariant derivative in
flat space. This theory is conformal, since the beta function of $g$
vanishes \cite{nonre}, but the anomalous dimension of $\psi $ is different
from zero. Precisely, (\ref{conf}) is a one-parameter family of conformal
field theories, parametrized by $g$. I\ use two-component complex spinors
and consider $n_{f}$ copies of them. The Dirac matrices are Hermitean and
such that $\gamma _{\mu }^{T}=-\gamma _{2}\gamma _{\mu }\gamma _{2}$, where $%
T$ means transpose. The ``time'' coordinate is $x_{3}$.

\bigskip

\textbf{Discrete symmetries.} The parity, charge-conjugation and
``time''-reversal transformations are 
\begin{eqnarray*}
P_{1} &:&\qquad x_{\mu }\rightarrow (-x_{1},x_{2},x_{3}),\qquad \psi
\rightarrow \gamma _{1}\psi ,\qquad \overline{\psi }\rightarrow -\overline{%
\psi }\gamma _{1},\qquad \\
&&\qquad \qquad \qquad \qquad \qquad \qquad \qquad A_{\mu }\rightarrow
(-A_{1},A_{2},A_{3})\qquad g^{2}\rightarrow -g^{2}. \\
C &:&\qquad x_{\mu }\rightarrow x_{\mu },\qquad \psi \rightarrow \gamma
_{2}\left( \overline{\psi }\right) ^{T},\qquad \overline{\psi }\rightarrow
-\psi ^{T}\gamma _{2},\qquad A_{\mu }\rightarrow -A_{\mu }\ . \\
T &:&\qquad x_{\mu }\rightarrow (x_{1},x_{2},-x_{3}),\qquad \psi \rightarrow
\gamma _{3}\psi ,\qquad \overline{\psi }\rightarrow -\overline{\psi }\gamma
_{3},\qquad \\
&&\qquad \qquad \qquad \qquad \qquad \qquad \qquad A_{\mu }\rightarrow
(A_{1},A_{2},-A_{3}),\qquad g^{2}\rightarrow -g^{2},
\end{eqnarray*}
where $\overline{\psi }=\psi ^{\dagger }\gamma _{3}$. Only C and CPT\ are
true symmetries, since P and T change the sign of $g^{2}$.

\bigskip

\textbf{Regularization.} The ordinary dimensional-regularization technique
is not convenient for the theory (\ref{conf}), because of the difficulties
related to the $\varepsilon $ tensor and the trace of an odd product of
gamma matrices. Some observations on this issue are collected in the
appendix. Nevertheless, for the purpose of computing divergent parts of
two-loop graphs, where only simple poles appear, it is consistent to ignore
this problem and work with the dimensional technique. This reduces the
effort and simplifies the algebra. Instead, it is necessary to use an
alternative regularization technique to prove properties valid to all orders
in the perturbative expansion. A standard choice is to modify the
gauge-field propagator with higher-derivative terms in a gauge-invariant way
and match the fermion loops with loops of Pauli-Villars fields. This can be
achieved with a regularized lagrangian 
\begin{equation}
\mathcal{L}_{\mathrm{B}}=\overline{\psi }_{\mathrm{B}}D\!\!\!\!\slash_{%
\mathrm{B}}\psi _{\mathrm{B}}+\frac{1}{2g_{\mathrm{B}}^{2}}\varepsilon ^{\mu
\nu \rho }F_{\mathrm{B}\mu \nu }\left( 1-\frac{\Box }{\Lambda ^{2}}\right)
A_{\mathrm{B}\rho },  \label{regla}
\end{equation}
and a regularized functional integration measure 
\begin{equation}
\lbrack \mathrm{d}\overline{\psi }][\mathrm{d}\psi ][\mathrm{d}%
A]\prod_{j}\det \left( D\!\!\!\!\slash_{\mathrm{B}}+M_{j}\right)
^{c_{j}},\qquad \sum_{j}c_{j}=-1,\qquad \sum_{j}c_{j}M_{j}^{p}=0,
\label{misa}
\end{equation}
where $p=1,2,\ldots $ and the $M_{j}$ have to tend to infinity. The
determinants in (\ref{misa}) come from integrating out the Pauli-Villars
fields. The superscripts B mean bare. Finally, I identify 
\begin{equation}
\sum_{j}c_{j}\ln M_{j}/\mu =-\ln \Lambda /\mu ,\qquad \sum_{j}c_{j}M_{j}\ln
M_{j}/\mu =b\Lambda ,  \label{identio}
\end{equation}
$b$ being an unspecified numerical factor. Here $\mu $ denotes the
renormalization scale, but the conditions (\ref{misa}) ensure that the
identifications (\ref{identio}) are $\mu $-independent, and therefore
consistent with renormalization-group invariance. It is also consistent to
set $b=0$, to kill the linear divergence by default.

The regularized gauge-fixing terms are 
\[
\mathcal{L}_{\mathrm{gf}}=\frac{1}{2\lambda g^{2}}(\partial _{\mu }A_{\mu
})\left( 1-\frac{\Box }{\Lambda ^{2}}\right) (\partial _{\nu }A_{\nu })+%
\overline{C}\Box \left( 1-\frac{\Box }{\Lambda ^{2}}\right) C.
\]
The ghosts decouple, as usual.

\begin{figure}[tbp]
\centerline{\epsfig{figure=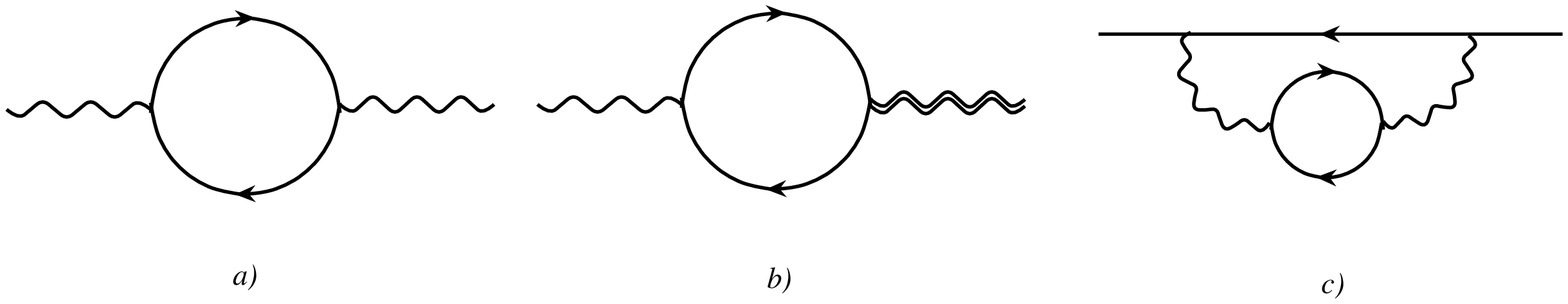,height=3cm,width=12cm}}
\caption{Simplest diagrams with the fermion bubble}
\label{fig1}
\end{figure}

\bigskip

\textbf{Renormalization.} The renormalized lagrangian reads 
\[
\mathcal{L}_{\mathrm{R}}=Z_{\psi }\overline{\psi }D\!\!\!\!\slash\psi +\frac{%
1}{2g^{2}}\varepsilon ^{\mu \nu \rho }F_{\mu \nu }\left( 1-\frac{\Box }{%
\Lambda ^{2}}\right) A_{\rho }+\Lambda \delta Z_{\Lambda }Z_{\psi }\overline{%
\psi }\psi 
\]
and the renormalization constants have expansions 
\[
Z_{\psi }=1+\sum_{n=1}^{\infty }a_{n}(g,\lambda )~(\ln \Lambda /\mu )^{n} 
\]
etc., where $\mu $ denotes the subtraction point, in the minimal subtraction
scheme. Standard Ward identities, combined with the properties of the
Chern-Simons term, ensure that the beta functions vanish, $\beta _{g}=\beta
_{\lambda }=0$ \cite{nonre}, and there is no need to insert renormalization
constants for the gauge field, $Z_{A}=Z_{g}=Z_{\lambda }=1$.

The perturbative results of this paper are written in the formalism of the
dimensional-regularization technique and are easily converted to the cut-off
regularization technique defined above replacing $1/\varepsilon $ with $\ln
\Lambda ^{2}/\mu ^{2}$ and understanding that power-like divergences are
subtracted in the conformal scheme (i.e. the scheme that preserves conformal
invariance at the quantum level).

The lowest-order values of the fermion renormalization constant and
anomalous dimension are given by the graph \textit{c}) of Fig. \ref{fig1}.
Up to subleading corrections in $1/n_{f}$ their values are 
\[
Z_{\psi }=1-\frac{g^{4}n_{f}}{384\pi ^{2}\varepsilon },\qquad \gamma _{\psi
}=\frac{1}{2}\frac{\mathrm{d}\ln Z_{\psi }}{\mathrm{d}\ln \mu }=\frac{%
g^{4}n_{f}}{384\pi ^{2}}. 
\]

\section{Gravity coupled with Chern-Simons $U(1)$ gauge theory and matter}

\label{gravicern} \setcounter{equation}{0}

The conventions for covariant derivatives $\mathcal{D}_{\mu }$, curvature $%
R^{a}$, torsion $\mathcal{D}e^{a}$ and spin connection $\omega _{\mu }^{a}$
are given in the appendix. The lagrangian of gravity coupled with
Chern-Simons $U(1)$ gauge theory and massless fermions reads 
\begin{equation}
\mathcal{L}=\frac{1}{2\kappa }eR+e\overline{\psi }\mathcal{D}\!\!\!\!\slash%
\psi +\frac{1}{2g^{2}}\varepsilon ^{\mu \nu \rho }F_{\mu \nu }A_{\rho },
\label{gravQED}
\end{equation}
where $e=\sqrt{g}$. The constant $\kappa $ has dimensionality $-1$ in units
of mass and serves as an expansion parameter for the irrelevant couplings.
The perturbative expansion is a double expansion in powers of $g$ and $%
\kappa E$, where $E$ is the energy scale.

The gravitational field is defined expanding the dreibein $e_{\mu }^{a}$
around flat space: 
\[
e_{\mu }^{a}=\delta _{\mu }^{a}+\phi _{\mu }^{a},\qquad \qquad \omega _{\mu
}^{a}=\varepsilon ^{abc}\partial ^{b}\phi _{\mu }^{c}+\mathcal{O}(\phi
^{2}), 
\]
and choosing the symmetric gauge $\phi _{\mu a}=\phi _{a\mu }$. It is
convenient to gauge-fix both gravity and the $U(1)$ gauge field in flat
space. The symmetric gauge is algebraic and so can be imposed directly. The
gauge-fixing sector of the theory is therefore 
\[
\mathcal{L}_{\mathrm{gf}}=\frac{1}{2\alpha \kappa }(\partial _{\mu }\phi
_{\mu \nu })^{2}+\frac{1}{2\lambda g^{2}}(\partial _{\mu }A_{\mu })^{2}+%
\mathcal{L}_{\mathrm{ghost}}. 
\]
The gauge parameters $\lambda $ and $\alpha $ are kept throughout the
calculations, because a powerful way to check the results is to check the
gauge-fixing independence of various quantities. The ghost part of the
gauge-fixing lagrangian is derived in detail in the appendix. The ghosts do
not contribute to the quantities calculated in this paper (this is proven in
section \ref{twolo}).

\bigskip

\textbf{Regularization.} The theory (\ref{gravQED}) is power-counting
non-renormalizable, and therefore, up to miraculous cancellations (that the
results of this paper exclude), divergences can be removed introducing
infinitely many new coupling constants, multiplying all possible irrelevant
operators. The vertices of the complete theory can contain arbitrarily many
derivatives and the regularized propagators should tend to zero faster than
any power at large momenta.\ The most convenient cut-off regularization
framework for the coupled theory is a Slavnov higher-derivative
regularization where propagators are exponentially corrected. For example,
the Chern-Simons field is regularized with 
\begin{equation}
\frac{1}{2g^{2}}\varepsilon ^{\mu \nu \rho }F_{\mu \nu }\exp \left( -\frac{%
\Box }{\Lambda ^{2}}\right) A_{\rho }+\text{non-minimal}.  \label{nonlo}
\end{equation}
The D'Alembertian is the covariant one and non-minimal terms have to be
fixed to ensure that the integrated regularized Chern-Simons term is gauge
invariant. To the order $1/\Lambda ^{2}$ we need to add 
\[
-\frac{1}{2g^{2}\Lambda ^{2}}\varepsilon ^{\mu \nu \rho }F_{\mu \nu }R_{\rho
\alpha }^{{}}A^{\alpha }. 
\]
It is immediate to prove that there exist appropriate non-minimal terms to
all orders in $1/\Lambda ^{2}$. Similar operations can be used to introduce
appropriate exponentials in the graviton and fermion propagators. However,
the exponentials regularize only the superficial divergences of diagrams
with more than one loop. One-loop divergences and subdivergences have to be
regularized apart, for example with the gauge-invariant Pauli-Villars method
of Fadeev and Slavnov \cite{slavnov}.

The existence of a manifestly gauge invariant regularization ensures the
absence of gauge anomalies to all orders in perturbation theory.

\bigskip

\textbf{Result.} The four-fermion divergent vertex (\ref{divgrav}) is
generated at the second order in the loop expansion and first order in the $%
\kappa $-expansion, up to subleading corrections in $1/n_{f}$. Therefore,
finiteness of three-dimensional gravity does not survive the coupling with
matter. To renormalize (\ref{divgrav}), it is first necessary to add new
vertices and coupling constants to the theory (\ref{gravQED}), 
\begin{equation}
\mathcal{L}=\frac{1}{2\kappa }eR+e\overline{\psi }\mathcal{D}\!\!\!\!\slash%
\psi +\frac{1}{2g^{2}}\varepsilon ^{\mu \nu \rho }F_{\mu \nu }A_{\rho
}+\kappa \lambda _{1}\frac{e}{4}(\overline{\psi }\psi )^{2}+\kappa \lambda
_{2}\frac{e}{4}(\overline{\psi }\gamma ^{a}\psi )^{2}+\mathcal{O}(\kappa
^{2}).  \label{Lprimo}
\end{equation}
Then, it is necessary to renormalize the new couplings by means of suitable
renormalization constants, $\lambda _{1,2\mathrm{B}}=\lambda _{1,2}Z_{1,2}$,
and redefine the fields. The field redefinitions have the form 
\[
A_{\mu \mathrm{B}}=A_{\mu }+\mathcal{O}(\kappa ),\qquad \psi _{\mathrm{B}%
}=Z_{\psi }^{1/2}\psi +\mathcal{O}(\kappa ),\qquad e_{\mu \mathrm{B}%
}^{a}=e_{\mu }^{a}+\mathcal{O}(\kappa )\text{.} 
\]
As in every non-renormalizable theory, the number of couplings is expected
to grow indefinitely with the order of the perturbative expansion in $\kappa 
$. Thus (\ref{Lprimo}), as a fundamental theory, is not physically
predictive. Of course, it is still predictive as an effective field theory.

\section{Absence of the Lorentz-Chern-Simons term}

\setcounter{equation}{0}

The theory (\ref{gravQED}) is parity violating. A priori, renormalization
might generate a Lorentz-Chern-Simons counterterm (\ref{gravChS}). Now I
prove that this does not happen. Basically, this term is finite and
therefore it is possible to set its renormalized coupling to zero
consistently.

The Lorentz-Chern-Simons term has dimensionality 3 in units of mass, so the
contributions to its renormalization must be $O(\kappa ^{0})$. Two types of
diagrams can contribute:\textit{\ i)} diagrams with internal gravitons and
gravitational ghosts; \textit{ii)} diagrams with no internal gravitons nor
gravitational ghosts.

By dimensional counting, the diagrams of type \textit{i)} can only be
one-loop. Indeed, higher loop diagrams with internal gravitons and/or
gravitational ghosts contribute to the renormalization of lagrangian terms
with dimensionality greater than 3. One-loop diagrams, on the other hand,
have no logarithmic divergence in three dimensions.

The diagrams of type \textit{ii)} can be studied in external gravity, using
properties of the trace anomaly. For concreteness, I consider the case of
Chern-Simons $U(1)$ gauge theory with massless fermions, but the
generalization of the proof is immediate.

It is useful to include the Lorentz-Chern-Simons term in the renormalized
lagrangian of the theory embedded in external gravity. In general, the
Lorentz-Chern-Simons term has to be multiplied by a coupling constant $\zeta 
$ plus a counterterm $\Delta _{\zeta }$, 
\[
\mathcal{L}_{\mathrm{R}}=eZ_{\psi }\overline{\psi }\mathcal{D}\!\!\!\!\slash%
\psi +\frac{1}{2g^{2}}\varepsilon ^{\mu \nu \rho }F_{\mu \nu }A_{\rho
}+(\zeta +\Delta _{\zeta })~\varepsilon ^{\mu \nu \rho }\left( \omega _{\mu
}^{a}\partial _{\nu }\omega _{\rho }^{a}+\frac{1}{3}\omega _{\mu }^{a}\omega
_{\nu }^{b}\omega _{\rho }^{c}\varepsilon ^{abc}\right) +\mathcal{L}_{%
\mathrm{gf}}. 
\]
The regularizing terms are not written explicitly. Obviously, $Z_{\psi }$
and $\Delta _{\zeta }$ depend only on $g$ and the gauge-fixing parameter $%
\lambda $, but not on $\zeta $, because the Lorentz-Chern-Simons term is
just the identity operator, in external gravity. Now I prove that $\Delta
_{\zeta }=0$.

The stress tensor is expressed as a functional derivative of the action, 
\begin{equation}
T_{\mu }^{\nu }(x)=\frac{e^{\nu a}(x)}{2e(x)}\frac{\delta S}{\delta e^{\mu
a}(x)}+\frac{e_{\mu }^{a}(x)g^{\nu \rho }(x)}{2e(x)}\frac{\delta S}{\delta
e^{\rho a}(x)}.  \label{stress}
\end{equation}
In the differentiation, the gauge-fixing and ghost terms can be ignored,
since they add gauge-exact contributions, which do not affect the physical
correlation functions. Because of the non-local regularization (\ref{nonlo}%
), the functional differentiation of (\ref{stress}) is involved. It is
convenient to focus the attention on the integrated stress tensor 
\[
\int e(x)T_{\mu }^{\nu }(x)~\mathrm{d}^{3}x, 
\]
and in particular the integrated trace 
\begin{equation}
\int e(x)\Theta (x)~\mathrm{d}^{3}x=\int e(x)T_{\mu }^{\mu }(x)~\mathrm{d}%
^{3}x.  \label{space}
\end{equation}
Inside (\ref{space}), the differentiation (\ref{stress}) simplifies
considerably. For example, it is possible to treat the spin connection,
Kristoffel symbols and curvatures as constants, because the functional
differentiation of objects such as $\partial _{\mu }g_{\nu \rho }$ and $%
\partial _{\mu }e_{\nu }^{a}$ produces total derivatives, which are killed
by the space-time integration of (\ref{space}). In practice, the operation (%
\ref{space}) reduces to a constant Weyl rescaling. Since the theory depends
on a unique scale, at the bare level, namely the cut-off $\Lambda $, the
result is easily proved to be 
\begin{equation}
\int e(x)\Theta (x)~\mathrm{d}^{3}x=-2\int eZ_{\psi }\overline{\psi }%
\mathcal{D}\!\!\!\!\slash\psi -\Lambda \left. \frac{\partial S}{\partial
\Lambda }\right| _{\mathrm{B}},  \label{bar}
\end{equation}
where the subscript means that the bare fields and coupling constants are
kept fixed in the $\Lambda $-differentiation. Since the renormalization
constants depend only on $\ln \Lambda /\mu $, the expression (\ref{bar}) can
be easily converted into a renormalized equivalent, 
\begin{equation}
\int e(x)\Theta (x)~\mathrm{d}^{3}x=-2\int eZ_{\psi }\overline{\psi }%
\mathcal{D}\!\!\!\!\slash\psi -\Lambda \frac{\partial S}{\partial \Lambda }%
-\mu \frac{\partial S}{\partial \mu }.  \label{tita}
\end{equation}
Now, consider a convergent correlation function 
\begin{equation}
G(x_{1}\cdots x_{n},y_{1}\cdots y_{m},z_{1}\cdots z_{m})=\langle A_{\mu
_{1}}(x_{1})\cdots A_{\mu _{n}}(x_{n})~\overline{\psi }(y_{1})\cdots 
\overline{\psi }(y_{m})~\psi (z_{1})\cdots \psi (z_{m})\rangle  \label{G}
\end{equation}
The $\Lambda \partial /\partial \Lambda $ derivative of a renormalized
correlation function is zero in the $\Lambda \rightarrow \infty $ limit, by
definition. Such derivative is equal to the insertion of $-\Lambda \partial
S/\partial \Lambda $, so the first term of (\ref{tita}) can be ignored when
the integral of $\Theta $ is inserted inside these correlation functions.
The result is 
\[
\int e(x)\Theta (x)~\mathrm{d}^{3}x=-2\int eZ_{\psi }\overline{\psi }%
\mathcal{D}\!\!\!\!\slash\psi -\mu \frac{\partial S}{\partial \mu }. 
\]
Since the theory is conformal in flat space, the couplings do not run, apart
possibly from $\zeta $, and so the partial $\ln \mu $ derivatives of the
renormalization constants can be replaced with total $\ln \mu $ derivatives,
up to terms proportional to the $\zeta $ beta function $\beta _{\zeta }=-%
\mathrm{d}\Delta _{\zeta }/\mathrm{d}\ln \mu $. The result is 
\[
\int e(x)\Theta (x)~\mathrm{d}^{3}x=-2(1+\gamma _{\psi })\int eZ_{\psi }%
\overline{\psi }\mathcal{D}\!\!\!\!\slash\psi +\beta _{\zeta }\int
~\varepsilon ^{\mu \nu \rho }\left( \omega _{\mu }^{a}\partial _{\nu }\omega
_{\rho }^{a}+\frac{1}{3}\omega _{\mu }^{a}\omega _{\nu }^{b}\omega _{\rho
}^{c}\varepsilon ^{abc}\right) . 
\]

The integral signs can be removed up to total derivatives. The only
ambiguity is a term $\partial _{\mu }(\overline{\psi }\gamma ^{\mu }\psi )$,
which however cannot appear, because it violates the symmetry under charge
conjugation (see the appendix).

The result is 
\begin{equation}
\Theta (x)=-2(1+\gamma _{\psi })[\mathrm{E}_{\psi }]+\beta _{\zeta
}e^{-1}~\varepsilon ^{\mu \nu \rho }\left( \omega _{\mu }^{a}\partial _{\nu
}\omega _{\rho }^{a}+\frac{1}{3}\omega _{\mu }^{a}\omega _{\nu }^{b}\omega
_{\rho }^{c}\varepsilon ^{abc}\right)  \label{teta}
\end{equation}
where 
\[
\lbrack \mathrm{E}_{\psi }]=\frac{1}{2}Z_{\psi }\overline{\psi }%
\overleftrightarrow{\mathcal{D}\!\!\!\!\slash}\psi =\frac{1}{2}e^{-1}\left( 
\overline{\psi }\frac{\delta _{l}S}{\delta \overline{\psi }}+\frac{\delta
_{r}S}{\delta \psi }\psi \right) 
\]
and $\delta _{l}$, $\delta _{r}$ denote the left and right functional
derivatives, respectively. The operator $[\mathrm{E}_{\psi }](x)$ is
proportional to the fermion field equation and therefore is finite. An
immediate proof is that inserting $[\mathrm{E}_{\psi }](x)$ in the
correlation function (\ref{G}) simply multiplies it by 
\begin{equation}
\frac{1}{2e(x)}\sum_{i=1}^{m}[\delta (x-y_{i})+\delta (x-z_{i})].
\label{oper}
\end{equation}
This result is standard and follows from a functional integration by parts.

Moreover, the second term of (\ref{teta}) should simply not be there,
because the unintegrated Lorentz-Chern-Simons term is not Lorentz invariant,
while $\Theta $ is. Therefore, $\zeta $ does not run: 
\begin{equation}
\beta _{\zeta }=0,\qquad \Delta _{\zeta }=\text{constant}.  \label{concla}
\end{equation}
The constant can be moved inside $\zeta $, so it is safe to write $\Delta
_{\zeta }=0$.

Having proved that the renormalized coupling $\zeta $ does not run, it is
meaningful to set it to zero. This means that the subtraction scheme can be
adapted in such a way that the Lorentz-Chern-Simons term is absent at each
order of the perturbative expansion. It is worth mentioning that if the
Lorentz-Chern-Simons term is treated within the minimal subtraction scheme
(or any generic scheme), finite contributions can survive and have to be
removed by hand. These facts have been recently confirmed by a number of
explicit two-loop computations \cite{landbenv}.

To conclude, there exists a modified subtraction scheme where the finite
part of the Lorentz-Chern-Simons term is identically zero. This is
important, because the theory with $\zeta \neq 0$, known as ``topologically
massive gravity'' \cite{jackiw}, is physically inequivalent to the theory
with $\zeta =0$. In the rest of the paper I focus on the theory with $\zeta
=0$. On the other hand, it is easy to prove that a small nonzero $\zeta $
does not change the two-loop results of the next sections and does not
affect the conclusion that quantum gravity coupled with matter, as it
stands, is not finite in three dimensions.

\bigskip

The arguments of this section are completely general and apply to every
theory of matter coupled with gravity. The generalization is straightforward.

Observe that Lorentz invariance is crucial in the derivation. The point is
that the unintegrated trace operator $\Theta (x)$ is Lorentz invariant, but
the unintegrated Lorentz-Chern-Simons term is not. A similar argument proves
that the beta function of the Chern-Simons coupling $g$ is zero \cite{nonre}%
: the gauge invariance of $\Theta $ and the gauge non-invariance of the
unintegrated $U(1)$ Chern-Simons term are not compatible with a running of $%
g $. Instead, the invariance under diffeomorphisms is not helpful in this
kind of reasonings, since the unintegrated operator $\Theta (x)$ is not
invariant under diffeomorphisms.

The second crucial point is the possibility to reduce to the theory in
external gravity. This is a lucky situation. The aguments based on the trace
of the energy momentum tensor cannot be applied if gravity is dynamical,
where the ``energy momentum tensor'' (by which I mean the derivative (\ref
{stress}) of the action with respect to the metric) vanishes identicaly
using the field equations. Other definitions of the stress tensor for
quantized gravity are more tricky to use.

Finally, it is known that pure gravity can be related to a Chern-Simons
theory in three dimensions \cite{witten}. However, Witten's arguments for
finiteness are based on the possibility to express the action in a form that
does not contain the inverse dreibein $e_{a}^{\mu }$, nor the inverse metric
tensor $g^{\mu \nu }$. This is impossible if gravity is coupled with
propagating matter.

These remarks explain why gravity coupled with matter can be not finite
despite the fact that there exists no graviton.

\section{Two-loop calculations}

\label{twolo} \setcounter{equation}{0}

In this section I describe the general setting of the two-loop computations.

\bigskip

\textbf{Four-fermion vertices.} I focus on the irrelevant terms of
dimensionality four in units of mass, which are 
\begin{equation}
e\overline{\psi }\mathcal{D}\!\!\!\!\slash^{2}\psi ,\qquad eF_{\mu \nu
}F^{\mu \nu },\qquad \varepsilon ^{\mu \nu \rho }e_{\rho }^{a}F_{\mu \nu }%
\overline{\psi }\gamma ^{a}\psi ,\qquad e(\overline{\psi }\psi )^{2},\qquad
e(\overline{\psi }\gamma ^{a}\psi )^{2}.  \label{dim4}
\end{equation}
Only the last two terms are independent, as I now prove.

Considering the presence of irrelevant terms, necessary for renormalization,
the most general field equations have the form 
\begin{eqnarray}
&&\mathcal{D}\!\!\!\!\slash\psi =\mathcal{O}(\kappa ),\qquad \qquad \qquad
\qquad \qquad F_{\mu \nu }+\frac{ig^{2}}{2}e\varepsilon _{\mu \nu \rho
}e^{\rho a}\overline{\psi }\gamma ^{a}\psi =\mathcal{O}(\kappa ),
\label{list} \\
&&\frac{1}{2\kappa }\left( R_{\mu \nu }-\frac{1}{2}g_{\mu \nu }R\right) +%
\frac{1}{8}e_{\mu }^{a}\overline{\psi }\gamma ^{a}\overleftrightarrow{%
\mathcal{D}}_{\nu }\psi +\frac{1}{8}e_{\nu }^{a}\overline{\psi }\gamma ^{a}%
\overleftrightarrow{\mathcal{D}}_{\mu }\psi -\frac{1}{4}g_{\mu \nu }%
\overline{\psi }\overleftrightarrow{\mathcal{D}\!\!\!\!\slash}\psi =\mathcal{%
O}(\kappa ).  \label{list5}
\end{eqnarray}
The first counterterm of (\ref{dim4}) vanishes using the fermion field
equation, up to higher orders in $\kappa $, so it can be removed by means of
a field redefinition. The second and third counterterms in (\ref{dim4}) are
equal to the forth of (\ref{dim4}) up to terms proportional to the field
equation of the gauge field (\ref{list}) and terms of dimensionality greater
than four. Finally, it is immediate to prove, using Fierz identities, that
the independent four-fermion vertices are precisely the ones listed in (\ref
{dim4}).

\bigskip

The second term in (\ref{dim4}) is an ordinary gauge-field kinetic term. It
has to be removed with a field redefinition of the form 
\begin{equation}
A_{\mu }\rightarrow A_{\mu }+a\kappa ~e\varepsilon _{\mu \nu \rho }F^{\nu
\rho }+b\kappa ~e_{\mu }^{a}\overline{\psi }\gamma ^{a}\psi ,  \label{redefa}
\end{equation}
where $a$ and $b$ are numerical coefficients. A theory with a propagating
gauge-field is physically inequivalent to (\ref{gravQED})-(\ref{Lprimo}).
Moreover, power-counting has to be reconsidered and the calculations have to
be repeated using the complete gauge field propagator. Here I stick to the
theory (\ref{gravQED})-(\ref{Lprimo}).

For the calculations of this paper, the use of field equations to simplify
the counterterms amounts in practice to the replacements 
\begin{equation}
R_{\mu \nu \rho \sigma }\rightarrow 0,\qquad F_{\mu \nu }\rightarrow -\frac{%
ig^{2}}{2}e\varepsilon _{\mu \nu \rho }e^{\rho a}\overline{\psi }\gamma
^{a}\psi ,\qquad D\!\!\!\!\slash\psi \rightarrow 0.  \label{feqrid}
\end{equation}

\bigskip

Collecting these observations, the two-loop counterterms have the form 
\begin{equation}
\mathcal{L}_{\text{counter}}^{grav}=c\frac{\kappa }{4}e(\overline{\psi }\psi
)^{2}+d\frac{\kappa }{4}e(\overline{\psi }\gamma ^{a}\psi )^{2}+\mathcal{O}%
(\kappa ^{2}),  \label{irrela}
\end{equation}
and the values of the numerical coefficients $c$ and $d$ have to be
determined with an explicit computation. Since the one-loop diagrams are
convergent in three dimensions, subdivergences are absent at two loops and
the divergent parts are simple poles $1/\varepsilon $ or simple logs $\ln
\Lambda ^{2}/\mu ^{2}$.

\bigskip

\textbf{Reduction of the number of diagrams.} Observe that the counterterms (%
\ref{irrela}) are necessarily polynomial in the number $n_{f}$ of fermions.
It is immediate to check the at the second loop order they are at most
linear in $n_{f}$. A quadratic contribution in $n_{f}$ would come from two
fermion loops. Two fermion loops can be connected only by a four fermion
vertex, otherwise the diagram is either not one-particle irreducible or not
two-loop. Then, however, the diagram factorizes into the product of two
one-loop diagrams, which are convergent in three dimensions.

The number of two-loop diagrams contributing to the four-fermion vertices is
high. It is convenient to concentrate on the contributions proportional to $%
n_{f}$, given by the diagrams that contain one fermion loop. Fermion loops
with two external legs are shown in Fig. \ref{fig1} and appear frequently as
subdiagrams of the two-loop diagrams. Fermion loops with three gauge-field
legs or one graviton leg and two gauge-field legs can also appear inside the
two-loop diagrams.

As anticipated in section \ref{gravicern}, the ghosts do not contribute to
the results of this paper. Indeed, the relevant Feynman diagrams do not have
external ghost legs (diagrams with external ghost legs affect only the
gauge-trivial sector of the theory). Diagrams with internal ghost legs
giving linear contributions in $n_{f}$ must have a ghost loop and a fermion
loop. Arguing as above, these diagrams factorize into the product of two
one-loop subdiagrams and therefore converge.

The one-loop self-energy of the gauge field is given by Fig. \ref{fig1}-a)
and is equal to 
\[
-\frac{n_{f}}{16}\frac{1}{(k^{2})^{(1+\varepsilon )/2}}(\delta _{\mu \nu
}k^{2}-k_{\mu }k_{\nu }). 
\]
The $\varepsilon $-dependence in the power of $k$ is kept, because it
affects the pole parts of the two-loop diagrams that contain the fermion
bubble as a subdiagram.

Obvious considerations based on spin conservation imply that the
graviton-gauge-field self-energy of Fig. \ref{fig1}-b) is identically zero.
This fact can be immediately checked with an explicit calculation.

\bigskip

\textbf{Calculations.} The divergent parts of the diagrams can be evaluated
with the techniques that follow. First, the diagrams are contracted with
external momenta, Dirac matrices, Kronecker tensors and $\varepsilon $
tensors in all possible ways, and traced in spinor indices. Then, the
results of these operations are differentiated a sufficient number of times
with respect to the external momenta, to arrive at dimensionless integrals,
and the external momenta are set to zero. Scalar products of internal
momenta in the numerators are converted into sums of squares, using for
example 
\[
p\cdot q=\frac{1}{2}\left[ p^{2}+q^{2}-(p-q)^{2}\right] .
\]
After a number of such algebraic manipulations, the calculation is reduced
to a set of integrals of the form 
\begin{equation}
\int \frac{\mathrm{d}^{D}p}{(2\pi )^{D}}\frac{\mathrm{d}^{D}q}{(2\pi )^{D}}%
\frac{1}{[p^{2}]^{a}[q^{2}]^{b}[(p-q)^{2}]^{c}}  \label{abc}
\end{equation}
where $D=3-\varepsilon $ and $a,b,c$ are integers such that $a+b+c=3$. It is
convenient to imagine that the fermions have a mass, to avoid IR divergences
at zero external momenta, and in some diagrams it is also useful to give
fictitious masses to the $U(1)$ field and the graviton.

The unique non-trivial contributions comes from the two-loop ``master''
integral 
\begin{equation}
\int \frac{\mathrm{d}^{D}p}{(2\pi )^{D}}\frac{\mathrm{d}^{D}q}{(2\pi )^{D}}%
\frac{1}{p^{2}q^{2}(p-q)^{2}}=\frac{1}{32\pi ^{2}\varepsilon }+\text{finite
part.}  \label{inte}
\end{equation}
The other integrals (\ref{abc}) are convergent. Indeed, if $a,b,c$ are not
all equal to one, then at least one of them is zero or negative, so there
are only two denominators. Integrals with two denominators factorize,
eventually after a translation, into the product of two integrals of the
form 
\[
\int \frac{\mathrm{d}^{D}p}{(2\pi )^{D}}\frac{p_{\mu _{1}}\cdots p_{\mu _{n}}%
}{[p^{2}]^{m}} 
\]
which are convergent. Also the integral (\ref{inte}) can be reduced to the
product of two integrals, using the technique of partial integration \cite
{ciakov} 
\[
0=\int \frac{\mathrm{d}^{D}p}{(2\pi )^{D}}\frac{\mathrm{d}^{D}q}{(2\pi )^{D}}%
\frac{\partial }{\partial p_{\mu }}\frac{p_{\mu }}{p^{2}q^{2}(p-q)^{2}}, 
\]
but this operation factorizes a $1/\varepsilon $.

The manipulations described so far are reversible, in the sense that it is
possible to reconstruct the structure of the divergent parts of the
diagrams, using the fact that they are local in the external momenta.

\begin{figure}[tbp]
\centerline{\epsfig{figure=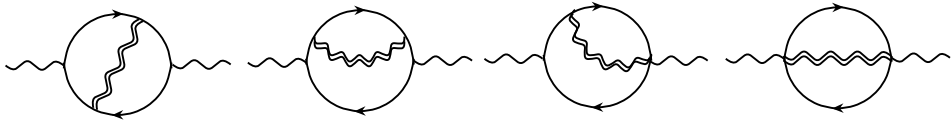,height=2cm,width=13cm}}
\caption{Two-loop self-energy of the gauge field with an internal graviton}
\label{fig3}
\end{figure}

\section{Four-fermion vertices induced by gravity}

\setcounter{equation}{0}

The counterterms of dimensionality 4 induced by gravity are proportional to
the Newton constant $\kappa $ and can therefore be computed at $\lambda
_{1}=\lambda _{2}=0$.

Three classes of diagrams contribute: the gauge-field two-point function,
the fermion-gauge-field three-point function and the fermion four-point
function. The divergent diagrams contain one internal graviton leg, one
fermionic loop and zero, one or two internal gauge-field legs, respectively.
There is no contribution to the fermion self-energy, since the potentially
relevant diagrams contain the subdiagram of Fig. \ref{fig1} \textit{b}). In
the figures, diagrams are depicted up to permutations of external legs and
reversions of the fermion arrows. It is not necessary to compute diagrams
with two external fermion legs plus two external gauge-field legs, because
they do not give new gauge-invariant contributions. To the order $\mathcal{O}%
(\kappa )$ it is not even necessary to consider diagrams with external
graviton legs. Such diagrams either contribute to the gauge-trivial sector
of the lagrangian or factorize a curvature tensor. Then, using (\ref{ricci})
and the graviton field equation of (\ref{list}), they can be converted into $%
\mathcal{O}(\kappa ^{2})$ counterterms.

\bigskip

\textbf{Gravitational contribution to the self-energy of the gauge field}.
The graphs that have an internal graviton leg and contribute to the two-loop
self-energy of the gauge field are shown in Fig. \ref{fig3}. The
counterterms associated with these graphs sum up to 
\begin{equation}
\mathcal{L}_{\text{counter-1}}=-\frac{5n_{f}\kappa \alpha }{384\pi
^{2}\varepsilon }eF_{\mu \nu }F^{\mu \nu }.  \label{Lcounter1}
\end{equation}
Observe that (\ref{Lcounter1}) is proportional to the gauge-fixing parameter 
$\alpha $ and should therefore be cancelled by some other contribution (see
below).

\bigskip

\begin{figure}[tbp]
\epsfig{figure=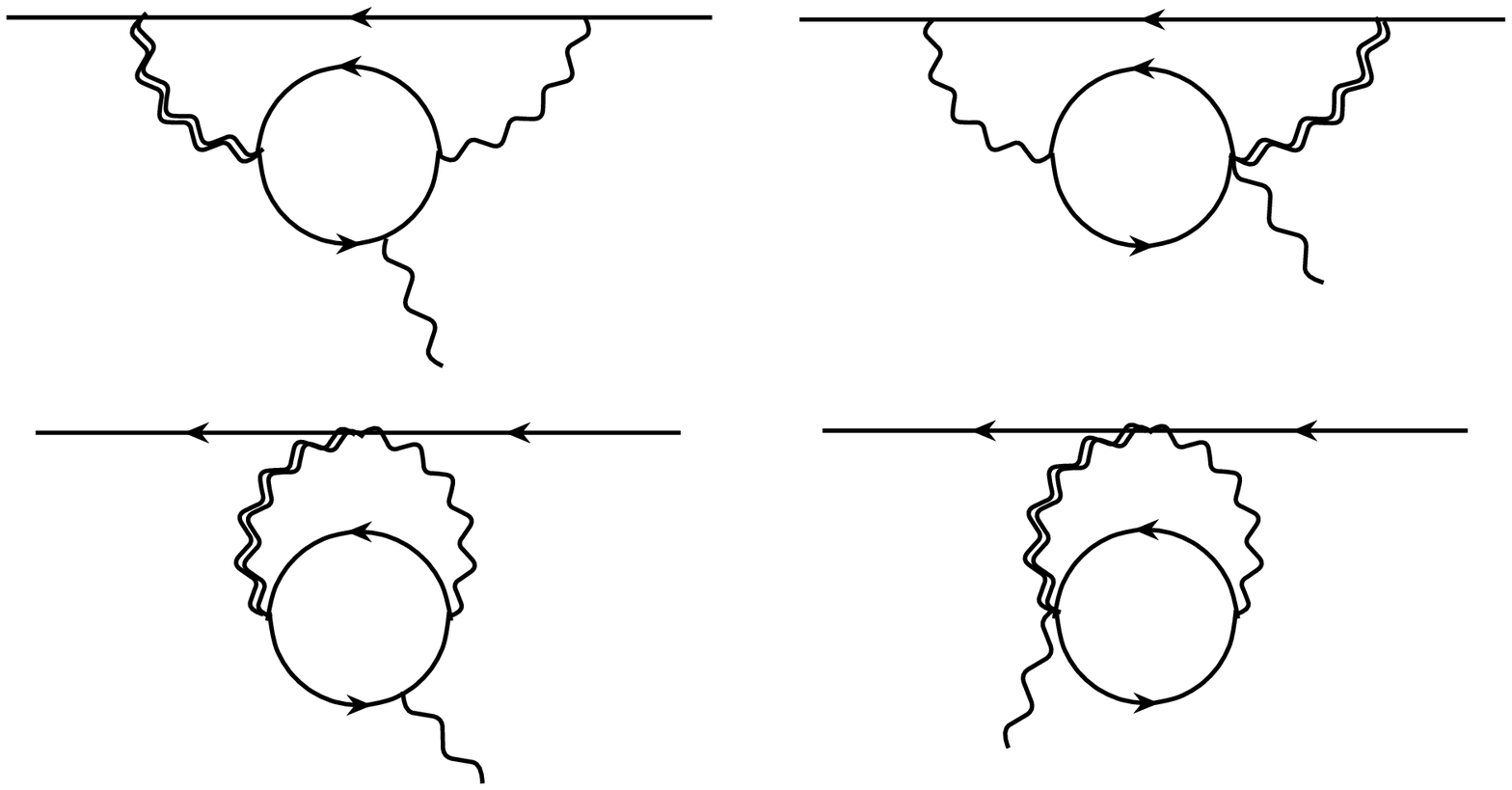,height=4cm,width=8cm} %
\epsfig{figure=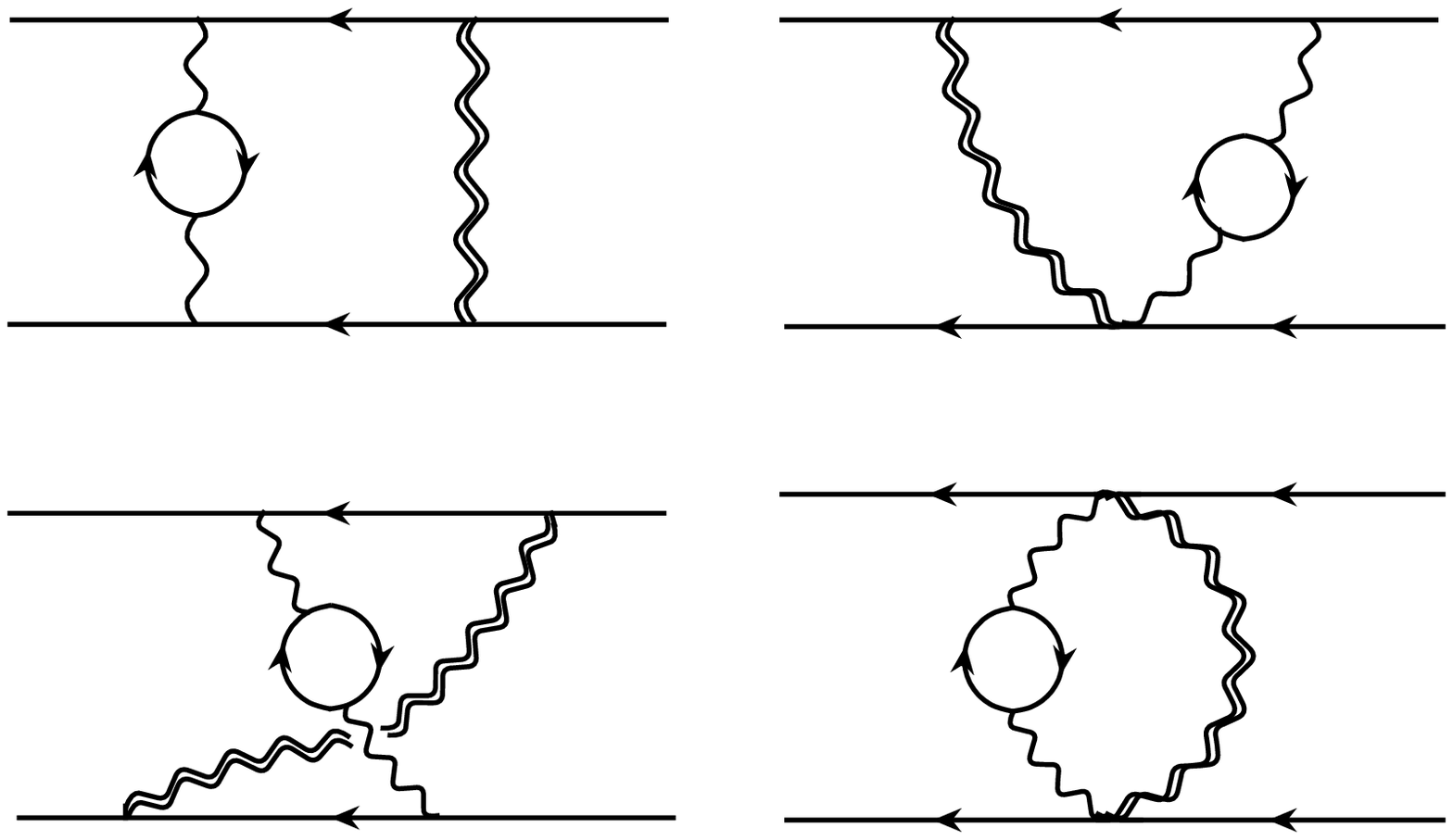,height=4cm,width=7.5cm}
\caption{Two-loop gravitational corrections to the fermion-gauge-field
vertex and to the four-fermion vertices}
\label{fig4}
\end{figure}

\textbf{Gravitational corrections to the fermion-gauge-field vertex.} The
divergent parts of these graphs, which are shown in the first half of Fig. 
\ref{fig4}, are subtracted by the counterterm 
\begin{equation}
\mathcal{L}_{\text{counter-2}}=-\frac{i\kappa g^{2}n_{f}}{768\pi
^{2}\varepsilon }\left( 3+10\alpha \right) \varepsilon ^{\mu \nu \rho
}e_{\rho }^{a}F_{\mu \nu }\overline{\psi }\gamma ^{a}\psi .
\label{Lcounter2}
\end{equation}
This is a Pauli term, but using the field equations (\ref{feqrid}) it can be
converted into a four-fermion vertex.

\bigskip

\textbf{Gravitational contribution to the fermion four-point function.} The
four-fermion counverterms that cancel the poles of the graphs shown in the
second half of Fig. \ref{fig4} are 
\begin{equation}
\mathcal{L}_{\text{counter-3}}=\frac{\kappa g^{4}n_{f}}{384\pi
^{2}\varepsilon }\left( 1+10\alpha \right) ~\frac{e}{4}(\overline{\psi }%
\gamma ^{a}\psi )^{2}.  \label{Lcounter3}
\end{equation}

Summing the three contributions (\ref{Lcounter1})-(\ref{Lcounter3}) and
using the substitutions (\ref{feqrid}), the $\alpha $-dependence drops out
and we obtain 
\[
\mathcal{L}_{\text{counter}}^{grav}=\mathcal{L}_{\text{counter-1}}+\mathcal{L%
}_{\text{counter-2}}+\mathcal{L}_{\text{counter-3}}=-\frac{5\kappa g^{4}n_{f}%
}{384\pi ^{2}\varepsilon }\frac{e}{4}(\overline{\psi }\gamma ^{a}\psi )^{2} 
\]
plus terms proportional to the field equations. The field redefinition that
reabsorbs the terms proportional to the field equations reads 
\[
A_{\mu }\rightarrow A_{\mu }-\frac{5n_{f}\alpha g^{2}\kappa }{768\pi
^{2}\varepsilon }~e\varepsilon _{\mu \nu \rho }F^{\nu \rho }-\frac{%
in_{f}g^{4}\kappa (3+5\alpha )}{768\pi ^{2}\varepsilon }~e_{\mu }^{a}%
\overline{\psi }\gamma ^{a}\psi . 
\]

\section{Conclusions}

\setcounter{equation}{0}

In this paper I have studied the renormalization of three-dimensional
quantum gravity coupled with matter. Using standard arguments it is possible
to show that the Lorentz-Chern-Simons term is not renormalized and therefore
there exists a subtraction scheme where it is identically absent. Instead,
irrelevant counterterms cannot be excluded \textit{a priori}. I have
performed a two-loop computation in a concrete model, gravity coupled with
Chern-Simons $U(1)$ gauge theory and massless fermions, to show that it is
not finite. The calculation can be simplified in various ways, but involves
a considerable number of diagrams. A good source of checks is the
gauge-fixing independence of the final result. A four-fermion counterterm 
\[
-\frac{5\kappa g^{4}n_{f}}{384\pi ^{2}\varepsilon }\frac{e}{4}(\overline{%
\psi }\gamma ^{a}\psi )^{2} 
\]
is turned on by renormalization. Therefore finiteness is violated at the
second order in the loop expansion and first order in the $\kappa $
expansion.

\bigskip

\textbf{Acknowledgement}. I am grateful to P. Menotti for drawing my
attention to references on 2+1 quantum gravity.

\section{Appendix}

\setcounter{equation}{0}

In this appendix I recall some basic formulas, useful to fix the notation
and simplify the analysis of the graphs. I also comment on the difficulties
to treat the Chern-Simons form in the context of the
dimensional-regularization technique.

\bigskip

\textbf{Curved-space conventions}. Torsion, curvatures, covariant
derivatives and connections are 
\begin{eqnarray*}
\mathcal{D}e^{a} &=&\mathrm{d}e^{a}+\varepsilon ^{abc}\omega
^{b}e^{c}=0,\qquad \qquad R^{a}=\mathrm{d}\omega ^{a}+\frac{1}{2}\varepsilon
^{abc}\omega ^{b}\omega ^{c}, \\
\mathcal{D}_{\mu }V_{\nu } &=&\partial _{\mu }V_{\nu }-\Gamma _{\mu \nu
}^{\rho }V_{\rho },\qquad \qquad \qquad \Gamma _{\mu \nu }^{\rho }=e^{\rho
a}\partial _{\mu }e_{\nu }^{a}+\varepsilon ^{abc}\omega _{\mu }^{a}e_{\nu
}^{b}e^{\rho c}, \\
\omega _{\mu }^{a} &=&\varepsilon ^{abc}\left( \partial _{\mu }e_{\nu
}^{b}-\partial _{\nu }e_{\mu }^{b}\right) e^{\nu c}-\frac{1}{4}e_{\mu
}^{a}\varepsilon ^{bcd}\left( \partial _{\rho }e_{\nu }^{b}-\partial _{\nu
}e_{\rho }^{b}\right) e^{\nu c}e^{\rho d}, \\
\mathcal{D}_{\mu }\psi &=&\partial _{\mu }\psi -\frac{i}{2}\omega _{\mu
}^{a}\gamma ^{a}\psi +iA_{\mu }\psi .
\end{eqnarray*}
The Ricci tensor and scalar curvature are defined as $R_{\mu \nu }=R_{\mu
\rho }^{ab}e^{\rho b}e_{\nu }^{a}$, $R=R_{\mu \nu }g^{\mu \nu },$ where 
\[
R^{ab}=\varepsilon ^{abc}R^{c}=\frac{1}{2}R_{\mu \nu }^{ab}\mathrm{d}x^{\mu }%
\mathrm{d}x^{\nu },\qquad {R^{\mu }}_{\nu \rho \sigma }=\partial _{\sigma
}\Gamma _{\nu \rho }^{\mu }-\partial _{\rho }\Gamma _{\nu \sigma }^{\mu
}-\Gamma _{\nu \sigma }^{\lambda }\Gamma _{\lambda \rho }^{\mu }+\Gamma
_{\nu \rho }^{\lambda }\Gamma _{\lambda \sigma }^{\mu }, 
\]
and of course $g_{\mu \nu }=e_{\mu }^{a}e_{\nu }^{a}$.

\bigskip

\textbf{Ghosts.} The ghosts are: $C$ for $U(1)$, $C^{\mu }$ for
diffeomorphisms and $C^{ab}=-C^{ba}$ for Lorentz rotations. Indices are
raised and lowered with $\delta _{\mu }^{a}$. The BRST variations of the
fields read 
\begin{eqnarray*}
sA_{\mu } &=&\partial _{\mu }C-A_{\rho }\partial _{\mu }C^{\rho }-C^{\rho
}\partial _{\rho }A_{\mu },\qquad \qquad se_{\mu }^{a}=-e_{\rho
}^{a}\partial _{\mu }C^{\rho }-C^{\rho }\partial _{\rho }e_{\mu
}^{a}-C^{ab}e_{\mu }^{b}, \\
sC &=&-C^{\rho }\partial _{\rho }C,\qquad sC^{ab}=-C^{ac}C^{cb}-C^{\rho
}\partial _{\rho }C^{ab},\qquad sC^{\rho }=-C^{\sigma }\partial _{\sigma
}C^{\rho }.
\end{eqnarray*}
It is necessary to introduce antighosts $\overline{C}$, $\overline{C}^{a}$
and $\overline{C}^{\mu a}=-\delta _{b}^{\mu }\delta _{\nu }^{a}\overline{C}%
^{\nu b}$. The ghost lagrangian reads 
\[
\mathcal{L}_{\mathrm{ghost}}=\partial _{\mu }\overline{C}(\partial _{\mu
}C-A_{\rho }\partial _{\mu }C^{\rho }-C^{\rho }\partial _{\rho }A_{\mu })+(%
\overline{C}^{\mu a}-\partial _{\mu }\overline{C}^{a})(e_{\rho }^{a}\partial
_{\mu }C^{\rho }+C^{\rho }\partial _{\rho }e_{\mu }^{a}+C^{ab}e_{\mu }^{b}). 
\]
Contracted indices may appear both as subscripts or superscripts in
Euclidean flat space. We have two ghost sectors: $U(1)$ and gravitational
(diffeomorphisms plus Lorentz symmetry). The two sectors have a diagonal
quadratic lagrangian, but mix due to a vertex of the form $\overline{C}%
AC^{\rho }$.

\bigskip

\textbf{Propagators.} The gauge-field propagator is 
\[
\langle A_{\mu }(k)~A_{\nu }(-k)\rangle _{\mathrm{free}}=-\frac{i}{2}%
g^{2}\varepsilon _{\mu \nu \rho }\frac{k_{\rho }}{k^{2}}+g^{2}\lambda \frac{%
k_{\mu }k_{\nu }}{k^{4}}. 
\]

The graviton propagator reads 
\begin{eqnarray*}
\langle \phi _{\mu \nu }(p)~\phi _{\rho \sigma }(-p)\rangle _{\mathrm{free}}
&=&\frac{\kappa }{2}\frac{1}{p^{2}}\left( \delta _{\mu \rho }\delta _{\nu
\sigma }+\delta _{\mu \sigma }\delta _{\nu \rho }-2\delta _{\mu \nu }\delta
_{\rho \sigma }\right) +\frac{\kappa }{p^{4}}\left( \delta _{\mu \nu
}p_{\rho }p_{\sigma }+p_{\mu }p_{\nu }\delta _{\rho \sigma }\right) + \\
&&\!\!\!\!\!\!\!\!\!\!\!\!\!\!\!{+\left( \alpha -\frac{1}{2}\right) \frac{%
\kappa }{p^{4}}\left( \delta _{\mu \rho }p_{\nu }p_{\sigma }+\delta _{\nu
\rho }p_{\mu }p_{\sigma }+\delta _{\mu \sigma }p_{\nu }p_{\rho }+\delta
_{\nu \sigma }p_{\mu }p_{\rho }\right) -3\alpha \kappa \frac{p_{\mu }p_{\nu
}p_{\rho }p_{\sigma }}{p^{6}}.}
\end{eqnarray*}

\bigskip

\textbf{Difficulties of the dimensional-regularization technique in curved
space.} Here I collect some observations about the difficulties to define a
consistent dimensional regularization for the Chern-Simons term in curved
space. If the theory contains two-component spinors, it is possible to
define the tensor 
\[
E^{abc}=-\frac{i}{2}\mathrm{tr}[\gamma ^{a}\gamma ^{b}\gamma ^{c}], 
\]
where $\gamma ^{a}$ are the dimensionally continued Pauli matrices. If the
trace is assumed to be cyclic, the $E$ tensor is set to zero by the
dimensional regularization \cite{collins}. However, if the theory contains
two-component fermions, the $D=3$ limit of $E^{abc}$ should be the ordinary $%
\varepsilon $ tensor. In curved space the situation worsens. Since the Pauli
matrices are constant and covariantly constant, so is the $E$ tensor,
assuming that it exists: $\partial _{\mu }E^{abc}=\mathcal{D}_{\mu
}E^{abc}=0 $. The Bianchi identity following from these equations is 
\begin{equation}
R_{\mu \nu }^{ad}E^{dbc}+R_{\mu \nu }^{cd}E^{dab}+R_{\mu \nu }^{bd}E^{dca}=0.
\label{bianchi}
\end{equation}

To define the propagator of the $U(1)$ gauge field, it would be useful to
have an ``inverse'' of the $E$ tensor, for example an \underline{$E$} tensor
satisfying 
\begin{equation}
E^{abc}\underline{E}_{mnp}=\frac{1}{3!}\delta _{mnp}^{abc}.  \label{inverse}
\end{equation}
However, contracting (\ref{bianchi}) with the \underline{$E$} tensor it is
immediate to obtain 
\begin{equation}
(D-3)R_{\mu \nu }^{ab}=0,  \label{diffi}
\end{equation}
which implies that either the dimension of spacetime is exactly equal to 3
or the spacetime is flat.

Moreover, an identity similar to (\ref{bianchi}) holds with $R_{\mu \nu
}^{ad}$ replaced by $\underline{E}^{adm}$. This follows immediately from the
definition (\ref{inverse}). Then, 
\[
(D-3)\underline{E}^{abc}=0. 
\]
This implies that the $\underline{E}$ tensor does not exist in $D$
dimensions.

An alternative approach is to split the $D$ dimensional spacetime into the
tensor product of a three-dimensional spacetime and a $(-\varepsilon )$%
-dimensional spacetime, as is commonly done in four dimensions to define the
matrix $\gamma _{5}$ and the tensor $\varepsilon _{\mu \nu \rho \sigma }$ 
\cite{collins}. Let $\mu ,\overline{\mu },\widetilde{\mu }$ denote the $D$%
-dimensional, three-dimensional and $(-\varepsilon )$-dimensional spacetime
indices, respectively. The kinetic lagrangian of the $U(1)$ field lives in
the three-dimensional subspace. The $(-\varepsilon )$-dimensional component $%
A_{\widetilde{\mu }}$ of the $U(1)$ gauge-field appears only in the Dirac
term and thus has no kinetic term. A way to treat a situation like this can
be found in ref. \cite{largeN}, using the large-$n_{f}$ expansion. The
missing kinetic term is provided by the fermion bubble, which is leading in
the large-$n_{f}$ expansion. Alternatively, it is possible to add $F_{\mu
\nu }^{2}$ multiplied by $1/\Lambda $, where $\Lambda $ is a further
cut-off, that is sent to infinity after $\varepsilon \rightarrow 0$.

\end{document}